\documentclass[preprint,aps]{revtex4}
\usepackage{amsfonts}

\usepackage{epsf}
\usepackage{anysize}
\marginsize{1 in}{1 in}{1in}{1.40in}

\usepackage{graphicx}
\usepackage{dcolumn}
\usepackage{bm}
\usepackage{amsmath}
\usepackage{amssymb}
\usepackage{mathrsfs}
\usepackage[latin1]{inputenc}
\usepackage{pstricks,pst-node,pst-text,pst-3d}
\usepackage[all]{xy}

\makeatletter
  \newcommand\figcaption{\def\@captype{figure}\caption}
  \newcommand\tabcaption{\def\@captype{table}\caption}
\makeatother

\newcommand{\dd}{{\rm d}}

\newcommand{\sd}{Schr\"{o}dinger }

\newcommand{\tr}{{\rm Tr}}

\begin{document}


\title{A Data-Driven Gradient Algorithm for High-Precision Quantum Control}

\author{Re-Bing Wu}
\affiliation{Department of Automation, Tsinghua University,
Beijing, 100084, China \\ Center for Quantum Information Science and Technology, BNRist,
Beijing, 100084, China}\email{rbwu@tsinghua.edu.cn}
\author{Bing Chu}
\affiliation{School of Electronic and Computer Science, University of Southampton, Southampton SO17 1BJ, UK}\email{b.chu@soton.ac.uk}
\author{David H. Owens}
\affiliation{Department of Automation, Zhengzhou University, Zhengzhou, 450001, China \\ Department of Automatic Control and Systems Engineering, The University of Sheffield, Mappin Street, Sheffield S1 3JD, UK}\email{d.h.owens@shef.ac.uk}
\author{Herschel Rabitz}
\affiliation{Department of Chemistry, Princeton University,
Princeton, NJ 08544, USA}\email{hrabitz@Princeton.Edu}

\date{\today}

\begin{abstract}
In the quest to achieve scalable quantum information processing technologies, gradient-based optimal control algorithms (e.g., GRAPE) are broadly used for implementing high-precision quantum gates, but their performance is often hindered by deterministic or random errors in the system model and the control electronics. In this paper, we show that GRAPE can be taught to be more effective by jointly learning from the design model and the experimental data obtained from process tomography. The resulting data-driven gradient optimization algorithm (d-GRAPE) can in principle correct all deterministic gate errors, with a mild efficiency loss. The d-GRAPE algorithm may become more powerful with broadband controls that involve a large number of control parameters, while other algorithms usually slow down due to the increased size of the search space. These advantages are demonstrated by simulating the implementation of a two-qubit CNOT gate.
\end{abstract}

\maketitle

\section{Introduction}
In practical quantum information processing, high-precision implementation of universal quantum gates (usually involving 1$\sim$3 qubits) is vital. Although the current control technology has been able to meet the minimum requirement for quantum error correction \cite{Bennett1996} (e.g., the 0.6-1$\%$ error threshold for surface codes has been reached in superconducting circuits \cite{Barends2014}, ion-traps \cite{Harty2014}, quantum-dots \cite{Veldhorst2014} and nitrogen-vacancy centers in diamond \cite{Rong2015}), the achievable precision still needs to be improved in order to reduce the resource overhead required for scalable quantum computation \cite{Devitt2013}.

Towards this ``last mile" target, an effective method for gate tuneup is to optimize the control pulses by following the gradient direction of the error function, which popularly has one form known as the GRAPE (GRadient Ascending Pulse Engineering) algorithm \cite{Khaneja2005}. When supplied with abundant control resources, the algorithm is highly efficient in that the optimization almost always quickly converges to a global optimal solution, owing to the underlying expectation of finding an attractive trap-free optimal control landscape
\cite{Rabitz2004,Wu2012,Russell2017}. The GRAPE algorithm is by nature {\it offline} (or {\it ex situ} \cite{Ferrie2015}) because the optimization is usually with respect to a design model identified from {prior} experiments, and no real online data is used during the optimization process. Thus, the systematic errors in the design model (e.g., the identified Hamiltonian and the pulse distortion by a waveform generater), as well as the {uncharacterized} random noises in the system and pulses, limit the control precision. Regarding these items, the designed control pulses should be immune to the systematic errors and be robust to the random noises.

Online (or {\it in situ}) learning can in principle correct for the systematic errors by iteratively calibrating the control pulses based on measurement outcomes. This {learning control concept} can be traced back to the early 1990s in the control of molecules by training ultrafast laser pulses \cite{Judson1992}, which has been successful in hundreds of physical and chemical experiments \cite{Brif2010}. In most applications, the control objective is with respect to a target state or the ensemble average of some quantum observable, where the control fields are updated by heuristic optimization algorithms such as a genetic algorithm \cite{Judson1992} or evolutionary strategy \cite{Roslund2009a}. Learning control for quantum gate tune-up is much more difficult {than the aforementioned applications}, because the full characterization of the control outcome requires process tomography that {needs} many more experiments to measure additional observables at high precision. In existing protocols, the extra data acquisition problem is usually bypassed via randomized benchmarking (RB)  \cite{Corcoles2013}, which is much easier for gate error verification without having to fully reconstruct the gate matrix. Several RB-based learning algorithms have been proposed, e.g., the Nelder-Mead algorithm was used in \cite{Egger2014} and \cite{Kelly2014}, with applications to superconducting qubits. To exploit the attractive trap-free control landscape \cite{Rabitz2004}, gradient-based (or greedy) algorithms were also introduced to accelerate the online optimization, where extra measurements (proportional to the number of control variables) need to be done to estimate the full or partial gradient from the data  \cite{Roslund2009,Li2017,Lu2017,Ferrie2015,Rol2017}.

The complexity of online learning control algorithms mainly depends on the total experimental costs, while the numerical calculations on a computer is usually negligible when only a few qubits are involved. In the existing algorithms, the overall cost can be very high due to the required many iterations (mainly for RB-based optimization \cite{Egger2014,Kelly2014,Ferrie2015}) or the expensive measurements in each iteration (mainly for gradient-based optimization \cite{Lu2017,Li2017,Roslund2009}).

To further reduce the total experimental cost, we find that the design model, which is often used for obtaining a good initial guess for the control pulse, can play a new role in accelerating the succeeding online learning calibration process. This opportunity arises because the design model contains valuable {\it a priori} knowledge about the experimental system, which is obtained from elaborately designed offline measurements. This motivation leads to the algorithm proposed in this paper, in which the design model is embedded into the data-driven learning procedure to synthesize the gradient vector also utilizing data from process tomography. The algorithm can effectively reduce the number of iterations by predicting the gradient descent or ascent direction for quantum gate tuneup, which compensates for the increased cost of tomography. Besides, under circumstances where broad-bandwidth controls are required for {noise suppression or high-speed gate operations}, the total experimental cost of our algorithm may be further reduced, {while the corresponding cost usually increases with other algorithms.} We refer to the method presented here as a data-driven type of GRAPE algorithm, or d-GRAPE for short. The remainder of the paper is organized as follows. The d-GRAPE algorithm is described in Section \ref{Sec:d-GRAPE}, whose effectiveness in correcting model error and control pulse distortion is demonstrated through simulations in Section \ref{Sec:SIM}. Finally, conclusions are presented in Section \ref{Sec:CONCLUSION}.

\section{The Data-driven gradient algorithm}\label{Sec:d-GRAPE}
In this section, we will present the basic procedure of the data-driven gradient algorithm.
\subsection{The quantum control model}
We assume that the quantum control system is closed and governed by the following \sd equation:
\begin{eqnarray}\label{Eq:real model}
  \dot{U}(t) & = & -i\left[H_0 + \sum_{k=1}^mu_k(t)H_k\right]U(t),
\end{eqnarray}
where $U(t)\in \mathbb{C}^{N\times N}$ represents the quantum gate operation on the states, with $U(0) = \mathbb{I}_N$, the identity matrix; and $u_k(t)\in \mathbb{R}$, $k=1.\cdots,m$, are the control fields imposed on the control system. The free Hamiltonian $H_0$ and the control Hamiltonians $H_k$'s are $N\times N$ Hermitian matrices that steer the unitary $U(t)$.

In practice, the above Hamiltonians are never precisely known. Thus, any numerical calculation has to be based on a design model:
\begin{eqnarray}\label{Eq:design model}
  \dot{U}_D(t) & = & -i\left[H_{D,0} + \sum_{k=1}^mv_k(t)H_{D,k}\right]U_D(t),
\end{eqnarray}
that can be accessed by a computer. The free and control Hamiltonians in the design model (\ref{Eq:design model}) can be very close to those in the actual system (\ref{Eq:real model}), but they are always imprecise {to some degree}. The control pulses $v_k(t)$ in the design model are often chosen as piecewise-constant pulses to facilitate numerical simulation on a digital computer and in some experimental situations. Note that the (designed) control pulses $v_k(t)$ are usually not identical with the actual pulses $u_k(t)$ applied to the system, because the control signal produced by an arbitrary waveform generator (AWG) is often distorted due to various factors including electronic limitations and transmission through the control line to the qubit. Such distorted signals have rising and falling edges or other unanticipated features, which are sometimes called quantum gate bleedthrough \cite{Kelly2014}. For example, the distortion can be modeled by a linear filter described as follows
\begin{equation}\label{Eq:Waveform generator}
u_k(t) = \mathcal{D}[v_k(t)]=\int_0^\infty h(t-\tau)v_k(\tau)\dd \tau,
\end{equation}
where $h(t)$ is the impulse response of the linear filter. The control pulse is distortion free only when $h(t)=\delta(t)$ is the Dirac function. In the following, we will show how to correct these errors by learning from online data.

\subsection{From GRAPE to d-GRAPE}
The goal of quantum gate tune-up is to find proper design control pulse sequences $\{v_k(t)\}$ such that the generated control $\{u_k(t)\}$ can lead the system propagator $U(T)$ as close as possible to a desired unitary matrix $U_f$. This can be achieved by minimizing the {\it infidelity} function \cite{Nielsen2000}
\begin{equation}
\mathcal{J} = \frac{1}{2N}\|U(T)-U_f\|^2,
\end{equation}
{where the norm is defined as $\|X\|=\sqrt{\tr(X^\dag X)}$.}

There are different ways of utilizing the gradient to optimize the control pulse. We illustrate the concept in the paper with the typical steepest descent algorithm that updates the control pulses in the following fashion:
\begin{equation}\label{Eq:GA}
v_k(t,\ell+1)=v_k(t,\ell) - \alpha{(\ell)}\cdot
g_k(t,\ell),
\end{equation}
 where $g_k(t,\ell)=\frac{\delta \mathcal{J}}{\delta v_k(t,\ell)}$ is the gradient in the $\ell$-th iteration and $\alpha(\ell)$ is the learning rate that is chosen as a sufficiently small positive real number. Taking $U(T)$ as an implicit function of $\{v_k(t)\}$ through (\ref{Eq:real model}) and (\ref{Eq:Waveform generator}), we have
\begin{eqnarray*}\label{Eq:g_R}
g_k(t,\ell) &=&\int_0^\infty \frac{\delta \mathcal{J}}{\delta u_k(t',\ell)}\frac{\delta u_k(t',\ell)}{\delta v_k(t,\ell)}\dd t'\\
&=&\int_0^\infty \langle \Delta(T,\ell)  ,H_k(t',\ell)\rangle \frac{\delta u_k(t',\ell)}{\delta v_k(t,\ell)}\dd t'
\end{eqnarray*}
where $H_{k}(t,\ell)=U^\dag(t,\ell) H_{k}U(t,\ell)$ and the inner product is defined as $\langle X,Y\rangle=\tr(X^\dagger Y)$. The error matrix is
$$\Delta(T,\ell)=\frac{1}{2i}\left[U_f^\dag U(T,\ell)-U^\dag(T,\ell)U_f \right].$$ {The variation term in the integral is induced by the distortion of the control pulses.} In the linear case exemplified in (\ref{Eq:Waveform generator}), we have
\begin{equation}
\frac{\delta u_k(t',\ell)}{\delta v_k(t,\ell)}=h(t'-t).
\end{equation}

Because the true gradient function (\ref{Eq:g_R}) can never be precisely evaluated due to the unavailability of the true model of the system, a practical operation is to ignore the pulse distortion and calculate the gradient in an {\it offline fashion}, as follows
\begin{equation}\label{Eq:g_OL}
g_k^{\rm OL}(t,\ell) = \langle \Delta_D(T,\ell) ,H_{D,k}(t,\ell)\rangle,
\end{equation}
where $H_{D,k}(t,\ell)=U_D^\dag(t,\ell)H_{D,k}U_D(t,\ell)$ and $$\Delta_D(T,\ell)=\frac{1}{2i}\left[U_f^\dag U_D(T,\ell)-U_D^\dag(T,\ell)U_f \right]$$ are both computed from the design model. Since the optimization is completely blind without checking the control performance with experimental data, the learning process along this gradient direction will be inevitably guided to a false solution that is optimal for the design model but not for the actual system.

\begin{figure}
\centering
\includegraphics[width=1\columnwidth]{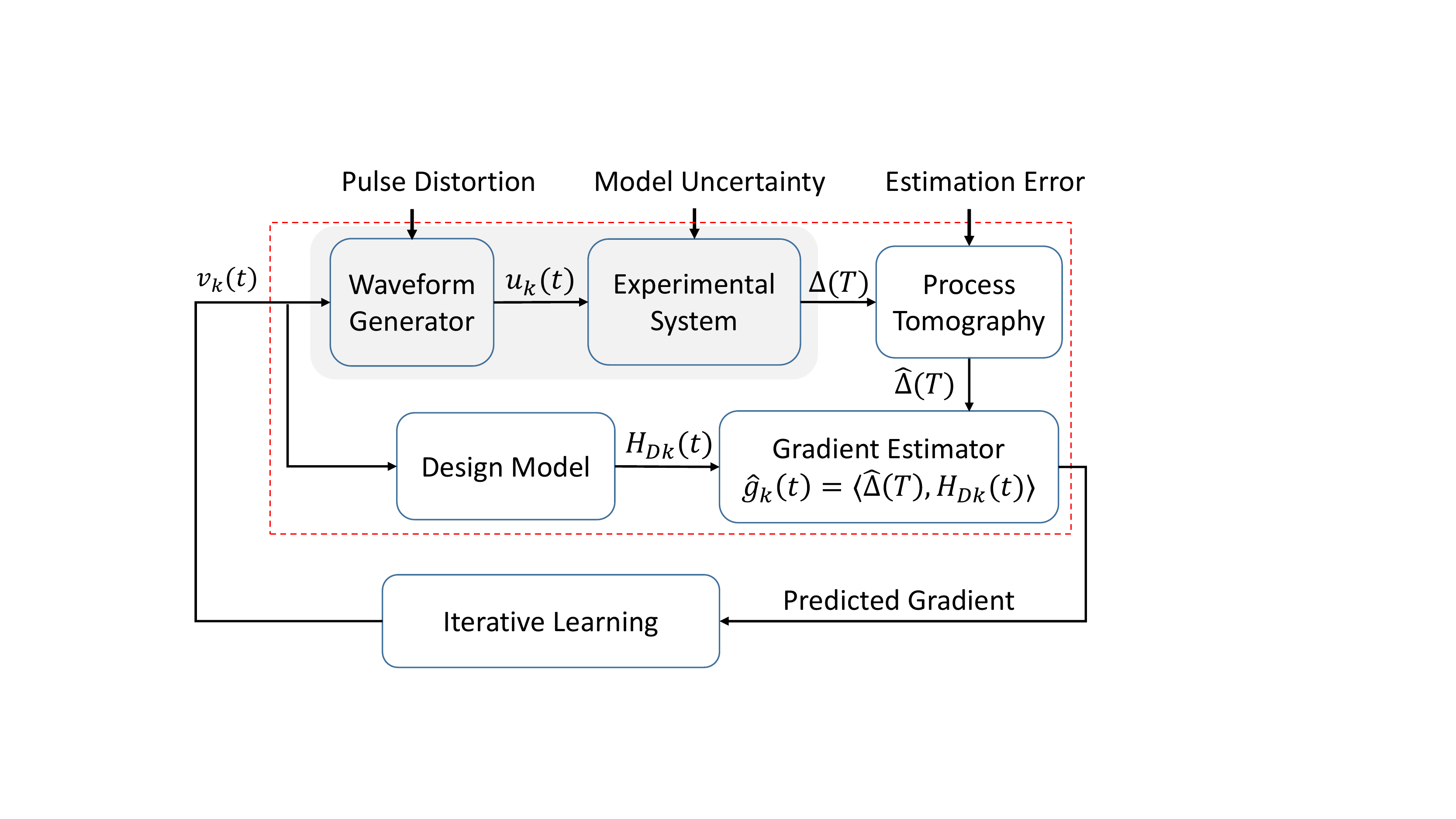}
\caption{(color online). Schematic diagram of the data-driven GRAPE (d-GRAPE) optimization procedure. The gradient is estimated from both the design model and the online data (for process tomography), which can in principle correct all deterministic errors such as the pulse distortion and the model uncertainty.} \label{fig:1}
\end{figure}

To find the genuine optimal control pulses, we take advantage of both of the above two approaches. The key concept is to estimate the gradient as follows:
\begin{equation}\label{Eq:g_MP}
\hat{g}_k(t,\ell) =\langle \hat{\Delta}(T,\ell) ,H_{D,k}(t,\ell)\rangle
\end{equation}
where the error matrix $\hat{\Delta}(T,\ell)$ comes form the estimation of $\Delta(T,\ell)$ through process tomography of $U(T)$, and $H_{D,k}(t,\ell)$ is calculated from the design model as in (\ref{Eq:g_OL}).
In this way, the real data are employed in order to deduce whether the learning algorithm is converging to a correct solution such that $\Delta(T,\ell)=0$, and its incorporation with $H_{D,k}(t,\ell)$ provides an approximate gradient whose deviation from the real gradient depends on the accuracy of the design model. The entire learning process is shown in Fig.~\ref{fig:1}, where the explicit use of the design model is the major difference with existing {\it model-free} learning control strategies in the literature. Therefore, we refer to the algorithm as d-GRAPE.

\subsection{Convergence analysis}
It is difficult to rigorously prove the convergence of the d-GRAPE to a globally optimal solution. Heuristically, d-GRAPE should converge to at least a locally optimal solution because the estimated gradient (\ref{Eq:g_MP}) can still maintain descent, although possibly not the steepest in presence of various uncertainties, as long as they are not too large. On the other hand, d-GRAPE can stop at a desired globally optimal control solution corresponding to $U(T)=U_f$ (assuming that the tomography error is negligible), where the gradient (\ref{Eq:g_MP}) vanishes. Therefore, when the system is controllable and the control resources are sufficiently abundant \cite{Wu2012}, the well-preserved attractive character of the control landscape should assure that d-GRAPE almost always converges to the desired global optimal solution, which will be verified in the following simulation examples. In principle, d-GRAPE is able to correct for any deterministic errors in the model or in the control pulses. Its precision is limited by that of the process tomography and other random noise sources in the system.

Compared with the existing online learning algorithms, d-GRAPE will be more competitive when broadband controls that involve a large number of variables are required for high precision, speed and robustness \cite{Zhang2015,Reiss2003}. Under such circumstance, the experimental cost of d-GRAPE per iteration will stay invariant, but the convergence may be faster owing to increased freedom in the control. However, the RB-based algorithms are expected to be more expensive because many more iterations are needed for search in the enlarged control space, as well as for the gradient-based algorithms proposed in \cite{Lu2017,Li2017,Roslund2009}, whose experimental costs per iteration increase with the number of control parameters.

\section{Simulations}\label{Sec:SIM}
In this section, we will show by numerical simulations how the algorithm can correct deterministic errors in the model and control pulses.

\subsection{Simulation Model}
We assume that the actual system is described by the following Hamiltonian:
\begin{eqnarray*}\label{}
H(t) &=& J\sigma_z^{1}\otimes\sigma_z^{2}+\sum_{i=1}^2\left[u_{x}^i(t)\sigma_x^{i}+u_{y}^i(t)\sigma_y^{i}\right],
\end{eqnarray*}
where $J$ is the coupling strength between the two qubits. The design model is as follows:
\begin{eqnarray*}\label{}
H_D(t) &=& (J+\delta J)\sigma_z^{1}\otimes\sigma_z^{2}+\sum_{i=1}^2\left[v_{x}^i(t)\sigma_x^{i}+v_{y}^i(t)\sigma_y^{i}\right],
\end{eqnarray*}
in which $\delta J$ represents the identification error of $J$ in the design model.

Moreover, we assume that the control pulses are distorted by a linear filter
$$u_{x,y}^i(t)=\int_0^t h(t-\tau)v_{x,y}^i(\tau)\dd \tau,\quad i=1,2,$$
in which the impulse response $h(t)$ is taken as
\begin{equation}
h(t) = \frac{1}{t_r}e^{-t/t_r},\quad t\geq 0.
\end{equation}
The the time constant $t_r$ characterizes the degree of pulse distortion by the steepness of the rising edge of distorted pulses. The pulses are heavily distorted when $t_r$ is long.

\subsection{Gate tuneup simulation results}
To demonstrate the ability of quantum gate tune-up by d-GRAPE, we test the target of a CNOT gate
$$U_f=\left(
        \begin{array}{cccc}
          1 & 0 &0  &0  \\
          0 & 1 & 0 & 0 \\
           0& 0 &0  & 1 \\
          0 & 0 & -1 & 0 \\
        \end{array}
      \right).
$$

{In the simulation, we set the coupling constant as $J=1$ and the final time as $T=5$. The time interval is evenly divided into $M=20$ sub-intervals, and hence the duration of each sub-interval is $\Delta t=T/M$.

\begin{figure}
\centering
\includegraphics[width=0.6\columnwidth]{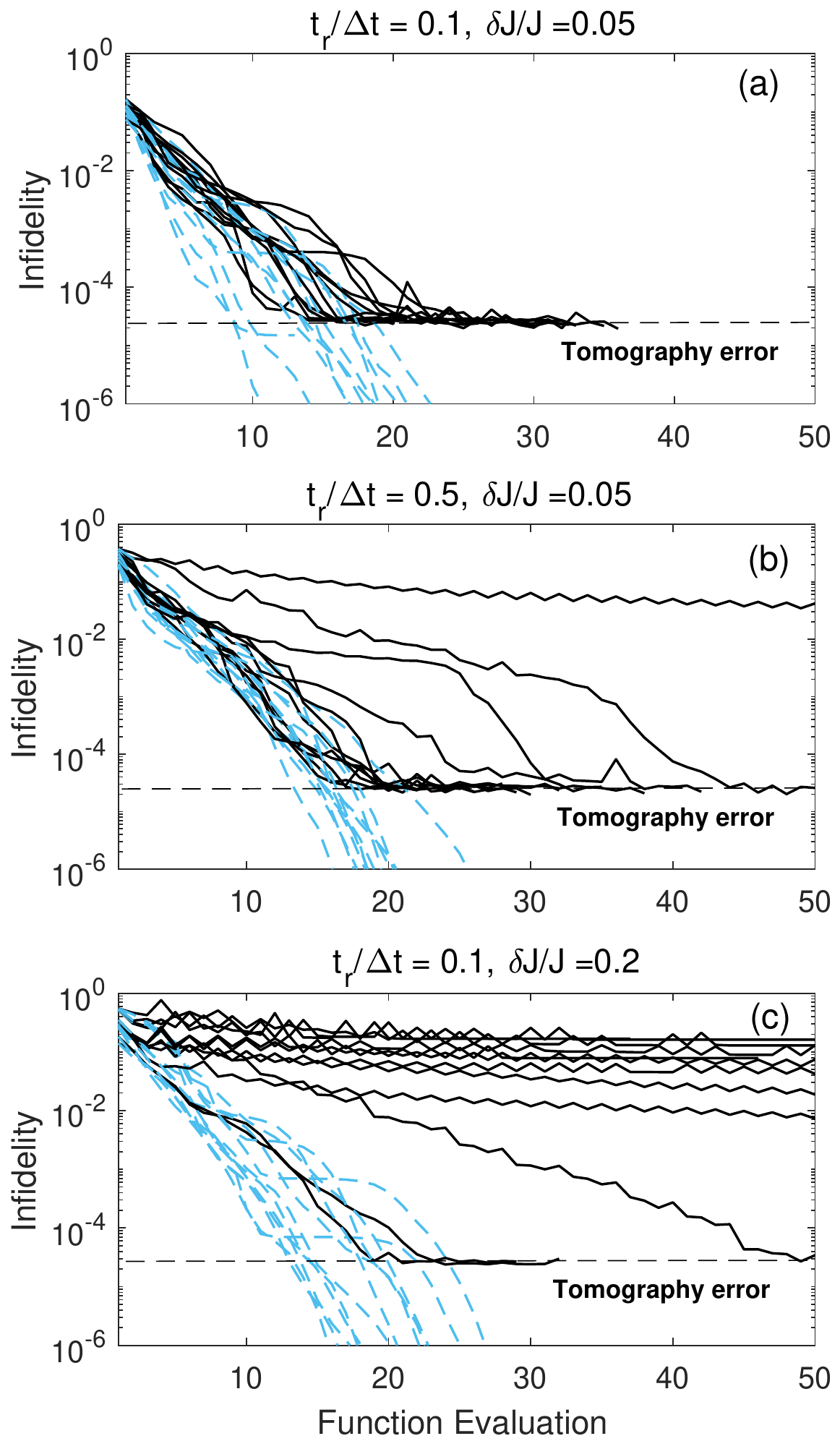}
\caption{(color online). Two-qubit quantum gate tuneup with the proposed d-GRAPE algorithm (solid black curves) for different model uncertainties and their comparison with ideal GRAPE algorithm (blue dashed curves). Each case include 12 runs from different initial guesses. (a) small pulse distortion $t_r/\Delta t = 0.1$ and small parametric error $\delta J/J=0.05$; (b) large pulse distortion $t_r/\Delta t = 0.5$ and small parametric error $\delta J/J=0.05$; (c) small pulse distortion $t_r/\Delta t = 0.1$ and large parametric error $\delta J/J=0.20$. The ultimate control precision $\sim 2\times 10^{-5}$ is limited by the estimation error of the process tomography (indicated by the horizontal dashed line).}\label{fig:2}
\end{figure}

\begin{figure}
\centering
\includegraphics[width=0.8\columnwidth]{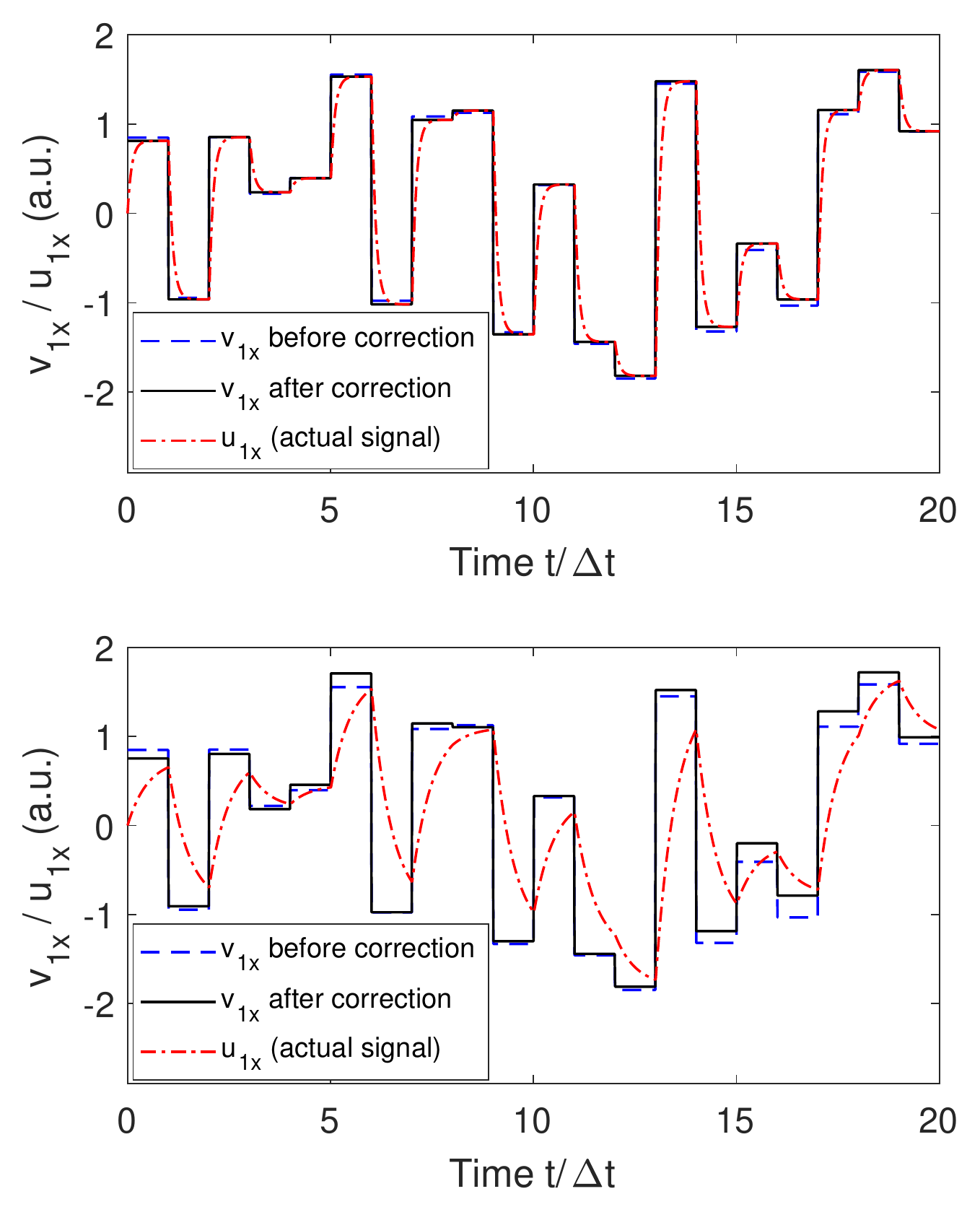}
\caption{(color online). The examples of optimized control pulses on the first qubit along the $x$-axis with $\delta J/J=0.1$. The pulse distortion parameters are $t_r/\Delta t=0.1$ (upper plot) and $t_r/\Delta t=0.5$ (lower plot). The blue dashed curves are the initial AWG reference signal, and the black solid curves are the corrected AWG signal optimized to the precision $\sim 2\times 10^{-5}$ that is limited by the tomography error. The actual distorted signals with rising and fall edges are shown by red dash-dotted curves. }\label{fig:3}
\end{figure}

In Fig.~\ref{fig:2}, we show three cases with parametric error $\delta J$ in $J$ and pulse distortion characterized by s$t_r$. Each case includes results from 12 different initial random guesses. We first offline optimize these fields [~i.e., following $g_k^{\rm OL}(t)$ in (\ref{Eq:g_OL})] to obtain a set of candidate pulses that are close to the optimal solution. Then, starting from these pulses, we perform d-GRAPE [~i.e., following $\hat{g}_k(t)$ in (\ref{Eq:g_MP})] based on the BFGS algorithm (a most popular gradient-based optimization algorithm \cite{Recipe2007}) that is more efficient than the steepest descent gradient algorithm. The estimation errors in the process tomography is simulated by injecting an additive random noise $\Delta U(\ell)$ (whose Frobenius norm is $\sim 2\times 10^{-5}$) to $U(T,\ell)$ in each iteration. For comparison, we also run the ideal GRAPE [~i.e., following $g_k(t)$ in (\ref{Eq:g_R}), assuming that both $\delta J$ and $t_r$ are precisely known] from the same set of initial pulses.

The simulation results show that the precision of the candidate pulses obtained from offline optimization (at the beginning of the optimization process shown in the plots) is always limited by the accuracy of the design model. When the model error is relatively small [see Fig.~\ref{fig:2}(a)], the succeeding optimization based on the proposed d-GRAPE algorithm (solid curves) almost always converges to its global optimal solution that is limited by the tomography error. Compared with the ideal GRAPE optimization (see the blue dash curves), its convergence speed is only slightly reduced.

When model error is not small enough (e.g., with severe pulse distortion in Fig.~\ref{fig:2}(b) or parameter deviation $\delta J$ in Fig.~\ref{fig:2}(c), fewer runs can quickly converge to the global optimal solution. Some runs still converge, but at the price of an increase in the number of iterations. We plot in Fig.~\ref{fig:3} the shapes of the corrected and actual $x$-axis control signals on the first qubit, showing that the d-GRAPE can correct for large pulse-distortion to achieve high-precision control without having to exactly know how the pulses are distorted.

The model error we choose in the simulations are relatively large (e.g, 20\% error in $J$ and pulse distortion $t_r/\Delta t=0.5$), and even under such a bad situation d-GRAPE can still tuneup the gate to some extent. When the model error gets larger, more and more optimizations become slower, and there even exist cases that the d-GRAPE algorithm gets lost and is trapped at a local false optimum solution. Thus, d-GRAPE should not be applied with a very coarse model, because of the potential traps and the increase experimental cost on process tomography. In practice, one should improve the precision of the design model as much as possible. Based on the high-precision model, the d-GRAPE algorithm can correct the error caused by the residue model imprecision within a few iterations.

\subsection{Comparison with RB-based algorithms}
We also tested the performance of d-GRAPE using different numbers of control pulses, and compared its performance with that of the gradient-free Nelder-Mead (NM) algorithm. The latter algorithm can be applied based on randomized benchmarking (RB) without having to use process tomography. The simulations are all based on a relatively precise model with $\delta J/J = 0.02$.

As shown in Fig.~\ref{fig:4}, when there are few pulses to tune ($M=10$), d-GRAPE is trapped over a very rugged control landscape (i.e., resulting from an insufficient number of control variables producing false landscape traps), while NM can struggle to achieve a high precision after about 2000 iterations. Given more control pulses, d-GRAPE can easily find high-precision control solutions over an almost trap-free control landscape in several tens of iterations. Correspondingly, the number of NM iterations is hundreds of times (or even over one thousand times) larger than that of d-GRAPE iterations. More importantly, the number of iterations increases for NM, but decreases for d-GRAPE, when the number of control pulses grows. Hence, d-GRAPE is supposed to outperform NM when the number of control variables is sufficiently large, which is expected to be the case {when higher precision and robustness are demanded.}

\begin{figure}
\centering
\includegraphics[width=1\columnwidth]{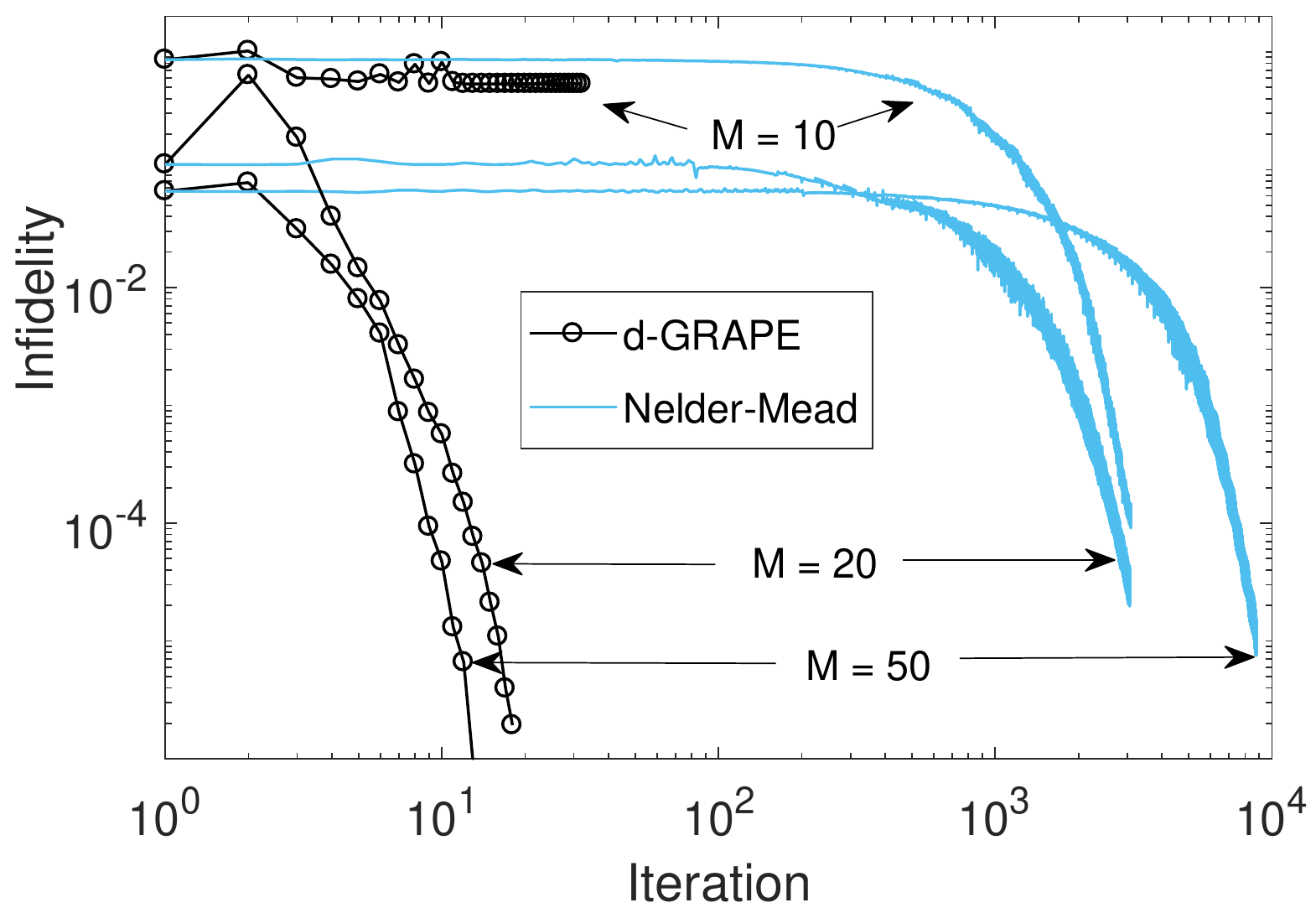}
\caption{(color online). Performance comparison of the d-GRAPE algorithm and the Nelder-Mead algorithm. {d-GRAPE may fail when there are only a few control parameters ($M=10$), and succeed with more control parameters ($M=20$ and $M=50$). When $M$ increases from 20 to 50, d-GRAPE becomes more efficient because less number of iterations are required, while the Nelder-Mead algorithm takes much more iterations to converge}. }\label{fig:4}
\end{figure}


\section{Conclusion and Discussion}\label{Sec:CONCLUSION}
To summarize, we have proposed a data-driven gradient (d-GRAPE) algorithm for optimizing laboratory control pulses against deterministic errors. The entire optimization procedure essentially performs both in a reinforcement learning manner from the online data in addition to supervised learning from the design model (or offline data). Analyses and simulations exemplify the calibration ability against errors induced by pulse distortion and model uncertainty, which is in principle extendable to more general non-uniform and nonlinear errors, as long as the process tomography can be done with sufficient precision and a reasonably good design model is available.

There is much room for the d-GRAPE algorithm to be improved. Several extensions of the algorithm are possible. First, extracting more knowledge from the offline model will improve the online optimization. For example, {we can estimate the gradient more precisely by incorporating the pulse distortion function $h(t)$ that can be offline identified from the waveform generator}; or we can use a more sophisticated learning algorithm such as a Newton algorithm, because the Hessian matrix can be estimated based on the same use of process tomography  without increasing the number of experiments. Second, combined with adaptive tomography \cite{Qi2017,Pogorelov2017,Straupe2016}, it is possible to simultaneously improve the precision of the control and the process tomography, which will further accelerate the learning process.

We also remark that d-GRAPE algorithm can be extended to more general objectives, e.g., quantum state preparation problems, where the cost of state tomography is cheaper and hence can be more efficient. When the real quantum system undergoes open dynamics, we can replace the unitary propagators by open-system process matrices, but the achievable precision may be limited by the decoherence effects. These potential topics and developments will be explored in the future.

\acknowledgments
The author RBW acknowledges support from NSFC grants (Nos. 61773232, 61374091 and 61134008) and National Key
Research and Development Program of China (Grant
No. 2017YFA0304300). The author HR acknowledges support from the ARO (W911NF-16-1-0014). The authors also acknowledge useful discussions with Professor Xinhua Peng and Dr. Xiaodong Yang from University of Science and Technology of China.


\end{document}